\documentclass{article}
\usepackage{graphicx} 
\usepackage{amsmath,amsthm,amssymb,amsfonts,amsxtra,amstext}
\usepackage{deluxetable}
\usepackage{hyperref}
\usepackage{authblk}

\usepackage[doi=false,isbn=false,url=false,style=numeric,bibstyle=numeric,sorting=nyt,maxbibnames=3,maxcitenames=2,backend=bibtex,firstinits=true]{biblatex}
\DeclareNameAlias{sortname}{last-first}
\usepackage{hyperref}
\title{Cross-Model Validation of Coronagraphic Exposure Time Calculators for the Habitable Worlds Observatory: A Report from the Exoplanet Science Yield sub-Working Group}

\author[a]{Christopher C. Stark}
\author[b]{Sarah Steiger}
\author[c]{Armen Tokadjian}
\author[d]{Dmitry Savransky}
\author[e]{Rus Belikov}
\author[c]{Pin Chen}
\author[c]{John Krist}
\author[f]{Bruce Macintosh}
\author[c]{Rhonda Morgan}
\author[b]{Laurent Pueyo}
\author[e]{Dan Sirbu}
\author[c]{Karl Stapelfeldt}

\affil[a]{NASA Goddard Space Flight Center, Greenbelt, MD 20771, USA}
\affil[b]{Space Telescope Science Institute, Baltimore, MD 21218, USA}
\affil[c]{Jet Propulsion Laboratory, California Institute of Technology, 4800 Oak Grove Drive, Pasadena, CA 91109, USA}
\affil[d]{Sibley School of Mechanical and Aerospace Engineering, Cornell University, Ithaca, NY 14853, USA}
\affil[e]{NASA Ames Research Center, Moffett Field, CA 94035, USA}
\affil[f]{Department of Astronomy \& Astrophysics, University of California, Santa Cruz, CA 95064, USA}

\begin{document}

\maketitle

\section{Introduction}

Estimating the exoplanet scientific productivity of the Habitable Worlds Observatory (HWO) requires estimating science exposure times\cite{brown2005}. From exoplanet yields to spectral retrievals, exposure times are at the heart of our understanding of the capabilities of this future mission. As such, ensuring accuracy and consistency between different exposure time calculators (ETCs) is critical.

For coronagraphic observations of potentially Earth-like planets, exposure time calculations require significant attention to detail. Many factors need to be considered when calculating exposure times of planets in the $10^{-10}$ contrast regime, from the effects of stellar diameter on the spatially-varying raw contrast, to stray light from binary companions, to the wavelength-dependent exozodi flux convolved with the coronagraph's spatially-varying off-axis PSF. Small differences in the treatment of these terms can sometimes lead to large differences in exposure times.

Various codes used by the community have independent coronagraphic ETCs at various levels of fidelity: the Altruistic Yield Optimizer (AYO) code\cite{stark2014}, EXOSIMS\cite{ExoSIMS}, the Error Budget Software (EBS)\cite{EBS}, the Planetary Spectrum Generator (PSG)\cite{checlair2021,kopparapu2021}, Bioverse\cite{bixel2021}, and others\cite{plavchan2024}. The Exoplanet Standards and Definitions Team previously led a calibration effort between AYO and EXOSIMS, with the goal of arriving at coarse agreement at the level of a factor of a few\cite{standardsteamreport}. Here we perform a detailed calibration effort designed to understand differences at the $\sim10\%$ level and smaller. 

This document summarizes the efforts of the Exoplanet Science Yield sub-Working Group's (ESYWG) ETC Calibration Task Group, which conducted a calibration study from March 4 to June 30 of 2024. Although many codes were considered for calibration efforts, three codes were ultimately calibrated as part of this study: AYO, EXOSIMS, and EBS. While EBS and EXOSIMS fundamentally use the same code base, they comprise separate branches of EXOSIMS; while they are not independent codes, some methods and assumptions differ, such as the base input target catalogs and the treatment of zodiacal light. AYO and EXOSIMS are almost entirely independent codes.

\section{Team}

The ETC Calibration Task Group was comprised of 14 individuals, experienced with coronagraphic yield and exposure time calculations, as well as coronagraph simulations and real-world observations. Christopher Stark led the Task Group and conducted all AYO exposure time calculations, Sarah Steiger conducted all EBS exposure time calculations, and Armen Tokadjian conducted all EXOSIMS exposure time calculations. Rhonda Morgan and Dmitry Savransky were consulted for EXOSIMS and EBS calculations. Dan Sirbu, Rus Belikov, and John Krist provided valuable insight into calculating stray light from nearby stars. 

The ETC Calibration Task Group was conducted in collaboration with the Coronagraph Design Survey (CDS; co-leads Rus Belikov and Christopher Stark) and the Coronagraph Technology Roadmap (CTR; co-leads Pin Chen and Laurent Pueyo), both of which were studies funded by the Exoplanet Exploration Program Office\cite{CDS,CTR}. 

Meetings of the ETC Calibration Task Group were held bi-weekly for three months. During these working meetings, the various parameters that feed into exposure time calculations were compared and checked, areas of disagreement were flagged, and sources of differences were determined.

\section{Summary of Assumptions}

Calibrating coronagraphic exposure times requires many assumptions about the performance of the instrument and spacecraft, as well as the astrophysical scene. Here we document the baseline assumptions made during this effort. These assumptions are not intended to accurately represent HWO. Rather, our assumptions were chosen to represent a reasonable point in design space that can act as a fiducial for comparison. As a result, we emphasize that absolute exposure times calculated during this effort likely do not reflect reality. This task group was focused on making sure the various count rates feeding into these times are in agreement. Once a more realistic design for HWO exists, absolute exposure times will reflect those of the mission. Where possible, this task group built off of the assumptions made during the CDS and CTR studies\cite{CDS,CTR}.

\begin{itemize}
\item Optical Telescope Element (OTE) assumptions
	\begin{itemize}
	\item An off-axis aperture defined by the Ultrastable Observatory Roadmap Team (USORT) study as our baseline\cite{USORT}--this segmented aperture consists of 19 hexagonal segments with an inscribed diameter of $6.5$ m and a circumscribed diameter of $7.87$ m.
	\item A Cassegrain telescope design with only two aluminum reflections in the coronagraphs' optical path.
	\end{itemize}
\item Coronagraph assumptions
	\begin{itemize}
    \item An optical layout similar to the LUVOIR study, with 15 silver surfaces in the coronagraph imaging channel\cite{LUVOIR} (end-to-end reflectivities were calculated using the same coating assumptions as the LUVOIR study\cite{LUVOIR} and are listed in Table \ref{table:missionassumptions}).
    \item A single visible wavelength coronagraph channel with 20\% bandwidth.
    \item Simultaneous observation of both polarizations without splitting on a single detector, with no polarization aberrations.
	\item Separate imaging and spectroscopy modes, with the spectroscopy mode using an integral field spectrograph (IFS) with the associated optical throughput reduction listed in Table \ref{table:missionassumptions}.
    \item An optical vortex coronagraph (OVC) designed for the USORT off-axis aperture, using the CDS-generated yield input package, accounting for raw contrast as a function of stellar diameter.
    \item Negligible jitter and perfect wavefront control.
	\end{itemize}
\item Detector assumptions
	\begin{itemize}
	\item Detector performance parameters (including QE and noise properties) identical to those of the HabEx and LUVOIR final reports\cite{HABEX,LUVOIR}.
	\end{itemize}	
\item Astrophysical assumptions
	\begin{itemize}
    \item All analyses adopt an Earth-twin at quadrature, at the Earth-equivalent insolation distance (EEID), where the EEID is proportional to the square root of the stellar bolometric luminosity. 
    \item A wavelength-independent planetary geometric albedo of 0.2.
    \item Three zodis of exozodiacal dust, where 1 zodi is 22 mags arcsec$^{-2}$ at V band and has an exozodi color that follows that of the target star.
    \item Local zodiacal flux calculated as a function of wavelength.
    \item Five fiducial stars defined by the CTR study, spanning F to late-K spectral types (HIPs 32439, 77052, 79672, 26779, and 113283).
    \end{itemize}  
\item Observational assumptions
	\begin{itemize}
    \item Exposure times calculated for broadband detections in bandpasses centered on 500 nm and 1000 nm.
    \item Exposure times calculated for spectral characterizations to detect O$_2$ at 800 nm and H$_2$O at 1000 nm.
    \item Simple circular aperture photometry using a radius of $0.7$ $\lambda/D_{\rm inscribed}$. 
	\item A systematic noise term in the denominator of the exposure time equation that causes exposure times to approach infinity as the signal count rate approaches that of the noise floor.
	\end{itemize}
\end{itemize}

\begin{deluxetable}{ccl}
\tablewidth{0pt}
\tabletypesize{\scriptsize}
\tablecaption{Baseline Astrophysical Parameters\label{table:astroassumptions}}
\tablehead{
\colhead{Parameter} & \colhead{Value} & \colhead{Description} \\
}
\startdata
$R_{\rm p}$ & $1.0$ $R_{\rm Earth}$ & Planet radius \\
$a$ & $1.0$ AU & Planet semi-major axis\tablenotemark{a} \\
$e$ & $0$ & Orbital eccentricity \\
$\theta$ & $\pi/2$ & Phase angle \\
$\Phi$ & Lambertian & Phase function \\ 
$A_G$ & $0.2$ & Planetary geometric albedo\\
$z$ & 23 mag arcsec$^{-2}$ & Average V band surface brightness of zodiacal light\tablenotemark{b} \\
$z'$ & 22 mag arcsec$^{-2}$  & V band surface brightness of 1 zodi of exozodiacal dust\tablenotemark{c} \\
$n$ & $3.0$ & Exozodi level of each star \\
\enddata
\vspace{-0.1in}
\tablenotetext{a}{For a solar twin.  This distance is scaled by $\sqrt{L_{\star}/L_{\odot}}$.}
\tablenotetext{b}{Varies with ecliptic latitude and solar longitude in some codes.}
\tablenotetext{c}{For Solar twin. Varies with spectral type and planet-star separation in some codes.}
\end{deluxetable}

\begin{deluxetable}{ccl}
\tablewidth{0pt}
\tabletypesize{\scriptsize}
\tablecaption{Coronagraph-based Mission Parameters\label{table:missionassumptions}}
\tablehead{
\colhead{Parameter} & \colhead{Value} & \colhead{Description} \\
}
\startdata
& & \bf{General Parameters} \\
$\tau_{\rm slew}$  & $1$ hr & Static overhead for slew and settling time \\
$\tau_{\rm WFC}$ & $1.3$ hr & Static overhead to dig dark hole \\
$\tau'_{\rm WFC}$ & $1.1$ & Multiplicative overhead to touch up dark hole \\
$D$ & $7.87$ m & Telescope circumscribed diameter (USORT aperture) \\
$D_{\rm ins}$ & $6.5$ m & Telescope inscribed diameter (USORT aperture) \\
$A$ & $427518$ cm$^2$ & Collecting area of telescope (USORT aperture) \\
$X$ & $0.7$ & Photometric aperture radius in $\lambda/D_{\rm ins}$ \\
$\Omega$ & $\pi(X \lambda/D_{\rm ins})^2$ radians & Solid angle subtended by photometric aperture \\
$\zeta_{\rm floor}$ & None & Raw contrast floor enforced regardless of coronagraph design \\
$T_{\rm contam}$ & $0.95$ & Effective throughput due to contamination applied to all observations \\

\hline
& & \bf{Detection Parameters for Bandpass 1} \\
$\lambda_{\rm d}$ & $0.5$ $\mu$m & Central wavelength for detection \\
$\Delta\lambda_{\rm d}$ & $20\%$ & Bandwidth assumed for detection \\
S/N$_{\rm d}$ & $7$ & S/N required for detection \\
$T_{\rm optical,d}$ & $0.56$ & End-to-end reflectivity/transmissivity at $\lambda_{\rm d}$ \\
$\theta_{\rm pix,d}$ & 6.55 mas & scale of detector pixel for detections \\

\hline
& & \bf{Detection Parameters for Bandpass 2} \\
$\lambda_{\rm d}$ & $1.0$ $\mu$m & Central wavelength for detection \\
$\Delta\lambda_{\rm d}$ & $20\%$ & Bandwidth assumed for detection \\
S/N$_{\rm d}$ & $7$ & S/N required for detection \\
$T_{\rm optical,d}$ & $0.56$ & End-to-end reflectivity/transmissivity at $\lambda_{\rm d}$ \\
$\theta_{\rm pix,d}$ & 6.55 mas & scale of detector pixel for detections \\

\hline
& & \bf{O$_2$ Characterization Parameters} \\
$\lambda_{\rm O2}$ & $0.8$ $\mu$m & Wavelength for characterization \\
S/N$_{\rm O2}$ & $10$ & Signal to noise per spectral bin evaluated in continuum \\
R$_{\rm O2}$ & 140 & Spectral resolving power \\
$T_{\rm optical,O2}$ & $0.32$ & End-to-end reflectivity/transmissivity at $\lambda_{\rm O2}$ including IFS optics \\
$n_{\rm pixperlenslet}$ & 6 & \# of pixels per lenslet per spectral bin in coronagraph IFS at $\lambda_{\rm O2}$ \\
$\theta_{\rm pix,c}$ & 6.55 mas & scale of detector pixel for characterizations \\

\hline
& & \bf{H$_2$O Characterization Parameters} \\
$\lambda_{\rm H2O}$ & $1.0$ $\mu$m & Wavelength for characterization \\
S/N$_{\rm H2O}$ & $5$ & Signal to noise per spectral bin evaluated in continuum \\
R$_{\rm H2O}$ & 140 & Spectral resolving power \\
$T_{\rm optical,H2O}$ & $0.32$ & End-to-end reflectivity/transmissivity at $\lambda_{\rm H2O}$ including IFS optics \\
$n_{\rm pixperlenslet}$ & 6 & \# of pixels per lenslet per spectral bin in coronagraph IFS at $\lambda_{\rm H2O}$ \\
$\theta_{\rm pix,c}$ & 6.55 mas & scale of detector pixel for characterizations \\

\hline
& & \bf{Detector Parameters} \\
$\xi$ & $3\times10^{-5}$ $e^-$ pix$^{-1}$ s$^{-1}$ & Dark current\\
RN & 0 $e^-$ pix$^{-1}$ read$^{-1}$ & Read noise\\
$\tau_{\rm read}$& 1000 s & Time between reads\\
CIC & $1.3\times10^{-3}$ $e^-$ pix$^{-1}$ frame$^{-1}$ & Clock induced charge\\
$T_{\rm QE}$ & $0.9$ & Raw QE of the detector at all wavelengths \\
$T_{\rm dQE}$ & $0.75$ & Effective throughput due to bad pixel/cosmic ray mitigation \\

\hline
\enddata
\vspace{-0.1in}
\label{coronagraph_params}
\end{deluxetable}

\section{Codes Benchmarked}

The ESYWG's ETC Calibration Task Group compared the exposure times for three codes: AYO, EXOSIMS, and EBS.

\subsection{AYO}

Exposure times for AYO are described in detail in \cite{stark2019}. Here we provide a brief overview, including several minor updates to \cite{stark2019}. 

AYO exposure times are calculated as
\begin{equation}
\label{eq:ayo_exposure_time}
	\tau = \left({\rm S/N}\right)^2 \left(\frac{{\rm CR_p} + 2\, {\rm CR_b} }{{\rm CR_p}^2 - {\rm S/N}^2 {\rm CR_{nf}}^2}\right),
\end{equation}
where ${\rm S/N}$ is the desired signal to noise ratio, ${\rm CR_p}$ is the photon count rate of the planet, ${\rm CR_b}$ is the photo count rate of all background sources, and ${\rm CR_{nf}}$ is the effective count rate of the systematic noise floor\cite{nemati2023}. The factor of two in the numerator doubling all background count rates is due to the assumption of ADI as a PSF-subtraction method. Equation \ref{eq:ayo_exposure_time} implicitly assumes that all backgrounds (zodiacal, exozodiacal, detector noise, stray light, etc.) can be subtracted down to the systematic noise floor.

The count rate of the planet is given by
\begin{equation}
\label{CRp_equation}
	{\rm CR_p} = F_0\, 10^{-0.4\left({\rm m}_{\lambda} + \Delta {\rm mag_p}\right)}\, A\, \Upsilon_{\rm c}\!\left(x,y\right)\, T\, \Delta \lambda,
\end{equation}
where $F_0$ is the zero-magnitude flux at the wavelength of interest $\lambda$, $m_{\lambda}$ is the stellar apparent magnitude at $\lambda$, $\Delta{\rm mag_p}$ is the magnitude difference between the planet and star, $\Delta\lambda$ is the bandwidth, $A$ is the effective collecting area of the telescope aperture accounting for segment gaps and secondary mirror/strut obscurations, $T$ is the total end-to-end optical throughput including detector quantum efficiency terms but excluding the coronagraph's core throughput, and $\Upsilon_{\rm c}\!\left(x,y\right)$ is the coronagraph's spatially-dependent core throughput. $\Upsilon_{\rm c}$ is defined as the fraction of light entering the coronagraph that ends up in the photometric core of the planet's PSF assuming perfectly reflecting/transmitting optics. By default the Altruistic Yield Optimizer's exposure time calculator optimizes the wavelength and photometric aperture used for signal extraction\cite{stark2024a,stark2024b}--for this effort, we turned these features off and prescribed a single wavelength and photometric aperture radius. 

The summed background count rate ${\rm CR_{b}} = {\rm CR_{b,\star}} + {\rm CR_{b,zodi}} + {\rm CR_{b,exozodi}} + {\rm CR_{b,detector}} + {\rm CR_{b,stray}}$. The leaked stellar count rate is given by
\begin{equation}
\label{CRbstar_equation}
	{\rm CR_{b,\star}} = F_0\, 10^{-0.4{\rm m}_{\lambda}}\, \frac{I\!\left(x,y\right)}{\theta^2} \Omega\, A\, T\, \Delta \lambda,
\end{equation}
where $I\!\left(x,y\right)/\theta^2$ is the spatially dependent leaked stellar count rate per unit solid angle exiting the coronagraph normalized to the starlight entering the coronagraph, and $\Omega$ is the solid angle of the photometric aperture used for planet detection.  

The zodiacal and exozodiacal background count rates are given by
\begin{equation}
\label{CRbzodi_equation}
	{\rm CR_{b,zodi}} = F_0\, 10^{-0.4z}\, \Omega\, A\, T\, T_{\rm sky}\!\left(x,y\right)\, \Delta \lambda,
\end{equation}
and
\begin{equation}
\label{CRbexozodi_equation}
	{\rm CR_{b,exozodi}} = F_0\, n\, 10^{-0.4z'\!\left(x,y\right)}\, \Omega\, A\, T\, T_{\rm sky}\!\left(x,y\right)\, \Delta \lambda,
\end{equation}
where $z$ is the surface brightness of the zodiacal light in magnitudes per unit solid angle, calculated at the desired wavelength and nominal pointing to the desired target, $z'$ is the surface brightness of 1 zodi of exozodiacal light in magnitudes per unit solid angle, and $n$ is the number of zodis assumed for all stars (details of AYO's treatment of exozodiacal surface brightness can be found in the appendices of \cite{stark2014}). $T_{\rm sky}\!\left(x,y\right)$ approximates the coronagraph's throughput for extended sources. Instead of convolving a 2D exozodi model with the coronagraph's spatially-dependent PSF, which is a numerically taxing process, we approximate the coronagraph's effects on our disk model by first convolving the coronagraph's spatially-dependent PSF at all locations with a normalized uniform background. AYO then simply multiplies each disk model by the resulting throughput factor, $T_{\rm sky}\!\left(x,y\right)$. 

The detector noise count rates are given by
\begin{equation}
 	{\rm CR}_{\rm b,detector} = n_{\rm pix} \left(\xi + {\rm RN}^2 / \tau_{\rm read} + 6.73\, {\rm CR}_{\rm sat}\, {\rm CIC}\right),
\end{equation}
where $n_{\rm pix}$ is the number of imager or integral field spectrograph (IFS) pixels per spectral element covered by the core of the planet's PSF, $\xi$ is the dark current, RN is the read noise, $\tau_{\rm read}$ is the length of an individual read, CIC is the clock induced charge, and CR$_{\rm sat}$ is the count rate of the brightest pixel for which we wish to achieve a given Geiger efficiency, nominally set to 10 times the brightest pixel in the image.

Stray light from binary companion stars is calculated via
\begin{equation}
\label{eq:AYO_stray_light}
	{\rm CR}_{\rm stray} = F_0\, 10^{-0.4\left({\rm m}_{\lambda} + \Delta {\rm mag}_{\rm b}\right)}\, \Omega\, A\, T_{\rm optical}\, T_{\rm sky}\!\left(x,y\right) P\left(x,y\right),
\end{equation}
where $\Delta {\rm mag}_{\rm b}$ is the difference in magnitude between the star and binary companion and $P\left(x,y\right)$ is the energy-normalized PSF per unit solid angle outside of the coronagraph's field of view. $P\left(x,y\right)$ is equal to $\zeta'(x,y)/\iint \zeta'(x',y') dx' dy',$ where $\zeta'$ is the normalized contrast of the PSF outside of the telescope's field of view. $\zeta'(x,y)$ is not well-known for HWO; we adopt a model of $\zeta'(x,y)$ developed for HabEx and provided by team member Dan Sirbu. Equation \ref{eq:AYO_stray_light} approximates the coronagraph mask transmission at point $(x,y)$ relative to the coronagraph center as $T_{\rm sky}\!\left(x,y\right)$. 

We note that Equation \ref{eq:AYO_stray_light} differs from that expressed in \cite{stark2019}. As part of this calibration effort we discovered an error in the equation used by \cite{stark2019}, which overestimated leaked stray light from binary stars by a factor of a few. More detail is provided on this in Section \ref{sec:code_updates}.

\subsection{EXOSIMS}

We model the signal as the sum of the photons from the planet ($c_p$) plus the photons from the background ($c_b$).  We define the rates of photon arrival from each of these sources (denoted by an overbar) such that our total measurement is given by
\begin{equation}
    z = \bar{c}_pt_\mathrm{int} +  \bar{c}_bt_\mathrm{int},
\end{equation}
where $t_\mathrm{int}$ is integration time.

The SNR metric is defined as
\begin{equation}
    \mathrm{SNR} = \frac{\bar{c}_pt_\mathrm{int}}{\sqrt{ \bar{c}_pt_\mathrm{int}  + \bar{c}_bt_\mathrm{int} + (\bar{c}_{sp}t_\mathrm{int})^2) }},
\end{equation}
where $\bar{c}_{sp}$ is the photon rate associated with the standard deviation of a noise component that fundamentally cannot be subtracted out from our measured signal (speckle residual). The integration time required to reach a target SNR is therefore
\begin{equation}
\label{eq:exosims_exposure_time}
    t_{\rm int} = \frac{ \bar{c}_p + \bar{c}_b }{\left(\dfrac{\bar{c}_p}{{\rm SNR}}\right)^2 - \bar{c}_{sp}^2},
\end{equation}
which imposes the condition
\begin{equation}
    \bar{c}_{sp} < \frac{ \bar{c}_p}{\mathrm{SNR}} \,.    
\end{equation}

In order to calculate exposure times in EXOSIMS, we use the built-in `calc\_intTime' function which is part of the OpticalSystem module. This function finds the integration time to reach a specified magnitude difference (dMag) at a particular working angle and zodi fluxes for a specific stellar system and using a specific observing mode. The following are the components needed to calculate the exposure time using this function:
\begin{itemize}
    \item TL (TargetList): We use the standard prototype TargetList module with slight modification (see below).
    \item sInds (indices of stars of interest): We set this to the indices pertaining to the following fiducial star list: HIP 32439, HIP 77052, HIP 79672, HIP 26779, HIP 113283
    \item fZ (local zodi flux): We use the Mennesson ZodiacalLight package which incorporates the zodi calculation outlined in \cite{stark2014}. Here, the location of the observatory is handled through longitude and latitude interpolation of the Leinert tables \cite{leinert1998} and then color corrected based on observing wavelength.
    \item fEZ (exozodi flux): We use the standard prototype ZodicalLight module, color-corrected according to local zodi (see Section 7.2.1).
    \item dMag (magnitude difference between planet and host star): Calculated using EXOSIMS util package and depends on planet albedo, radius, phase (set to quadrature), and star-planet distance.
    \item WA (working angle): Set specifically for each planet at quadrature using distance to planet from Earth and star-planet separation
    \item mode (observing mode): Holds information for specific instrument and whether it is detecting (imager) or characterizing (spectrograph), and set to the instrument parameters described in Table~\ref{coronagraph_params}.
\end{itemize}

There were some specific changes needed to run calculations for this calibration effort. We created a new SimulatedUniverse module to include exactly 1 planet for every 1 star in our StarCatalog (EXOCAT1\cite{turnbull2015}). We also modifed the TargetList module to skip the completeness filter so that the stars of interest will assuredly remain in the target list. With these modifications, the results presented here can be reproduced by simply importing the necessary modules from EXOSIMS and running the `calc\_inttime' method.

\subsection{EBS}

The Error Budget Software (EBS; \cite{EBS}) is a wrapper for EXOSIMS which was created to fold in wavefront error, wavefront sensing and control, and telescope stability information into EXOSIMS. This is achieved through calculating and changing the ``ppFact'' variable in EXOSIMS. ppFact is a post-processing factor that gets multiplied by $\bar{c}_{sp}$ in Equation \ref{eq:exosims_exposure_time} and essentially modifies the noise floor. 

For the purposes of this work, this functionality was suppressed by fixing ppFact to a single value and so EBS should behave identically to EXOSIMS with respect to all calculations. This does \textit{not} mean that EBS and EXOSIMS in this analysis will have identically the same calculated count rates and final exposure times. Several inputs to each of these codes differ, including the input star catalogs and methods used for zodiacal light calculations. 

EBS used the HWOMissionStars StarCatalog which pulls from the NASA ExEP HWO Target List\cite{exeptargetlist}. A modification had to be made for this study, however, to add fields to that catalog that track the separation and magnitude of any nearby stars in the case that the target was part of a multi-star system (as is the case for HIP 77052). This enters into the calculation in the form of a stray light noise term encompassed within $\bar c_{b}$. This data came in the form of a CSV file that was given to the ``wdsfilepath'' field of the EXOSIMS input JSON file. EBS also assumes a constant value for zodiacal surface brightness (23 magnitudes/arcsec$^2$) irregardless of target as it does not account for any changes due to telescope pointing. 

These input differences can easily lead to discrepant outputs providing useful insights as to their origin which will be explored further in later sections. 

\section{Calibration Method\label{sec:cal_method}}

Coronagraphic exposure times require detailed calculations using a large number of inputs. It is therefore not unexpected for exposure times to differ. Our Task Group set out to understand the fundamental sources of any differences that were discovered. As such, simply reporting exposure times, or even count rates, would be inadequate.

\begin{figure}
    \centering
    \includegraphics[width=1.0\linewidth]{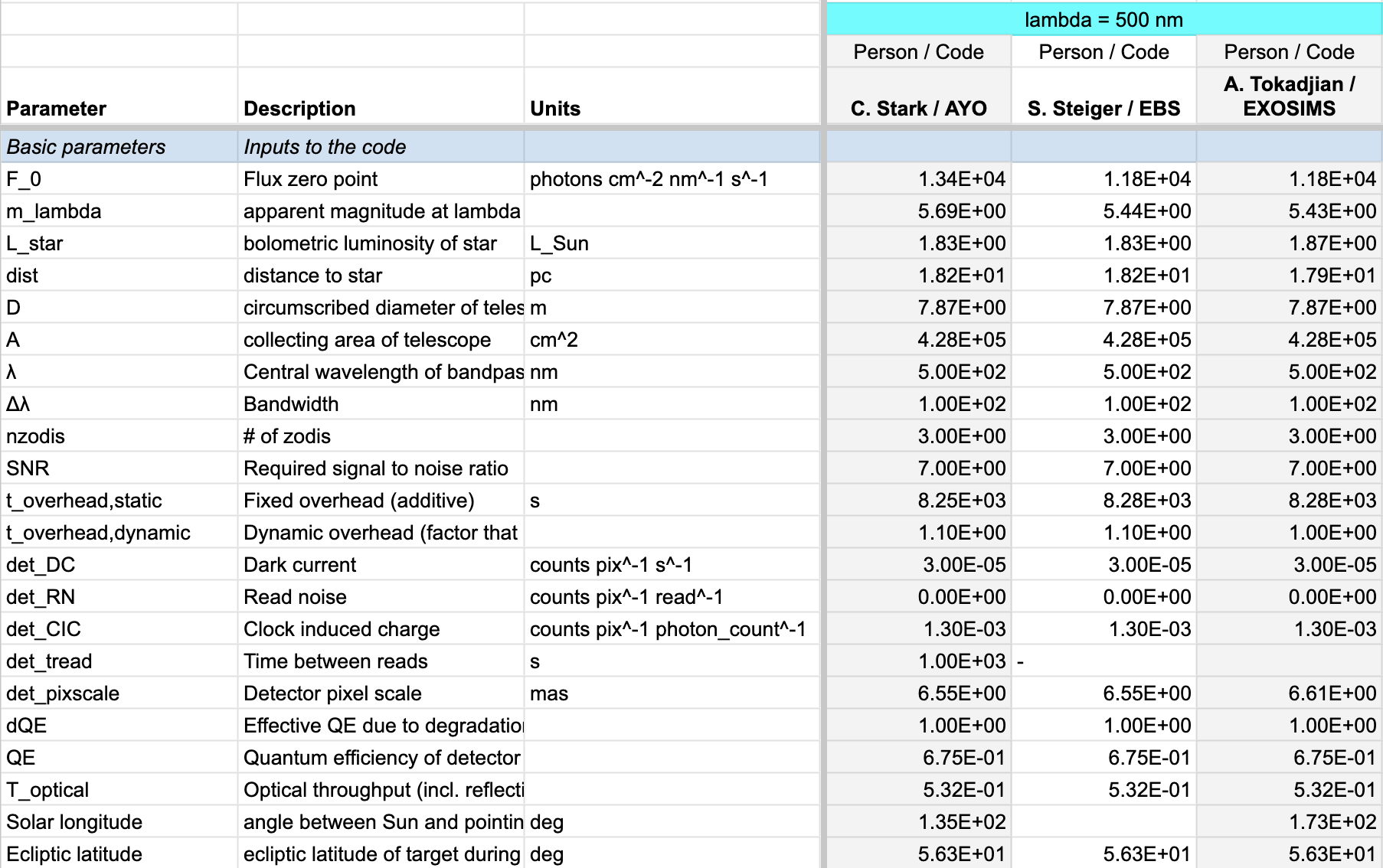}
    \caption{Example portion of the spreadsheets used for ETC calibration. This example spreadsheet corresponds to 500 nm detection observations around the star HIP 32439. Shown are the input parameters used for ETC calculations. Differences in the input target list lead to differences in several parameters that propagate through calculations.}
    \label{fig:spreadsheet_basic}
\end{figure}

\begin{figure}
    \centering
    \includegraphics[width=1.0\linewidth]{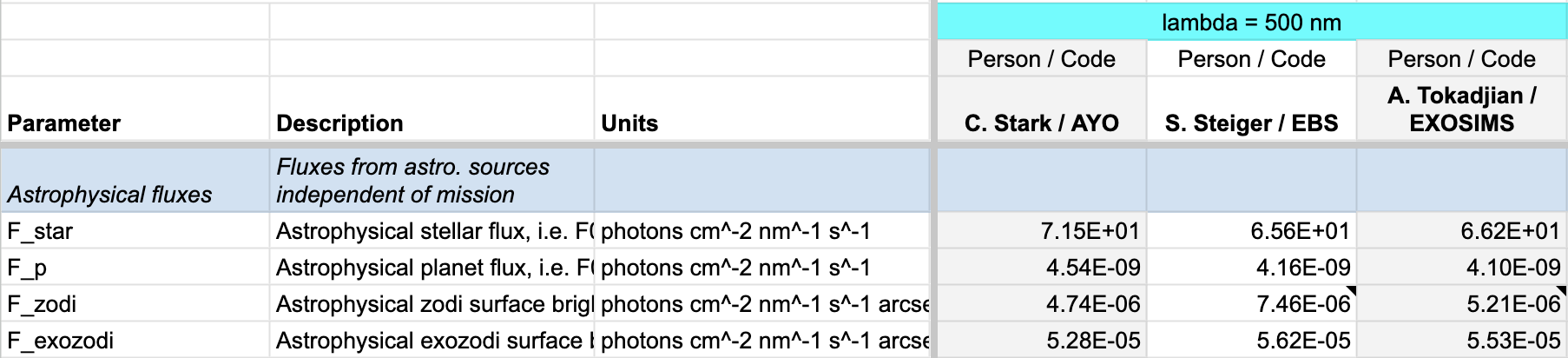}
    \caption{Same as Figure \ref{fig:spreadsheet_basic}, but listing the calculated astrophysical fluxes at the primary mirror. The causes of all differences in astrophysical fluxes were determined and code updates were made where appropriate.}
    \label{fig:spreadsheet_astro}
\end{figure}

\begin{figure}
    \centering
    \includegraphics[width=1.0\linewidth]{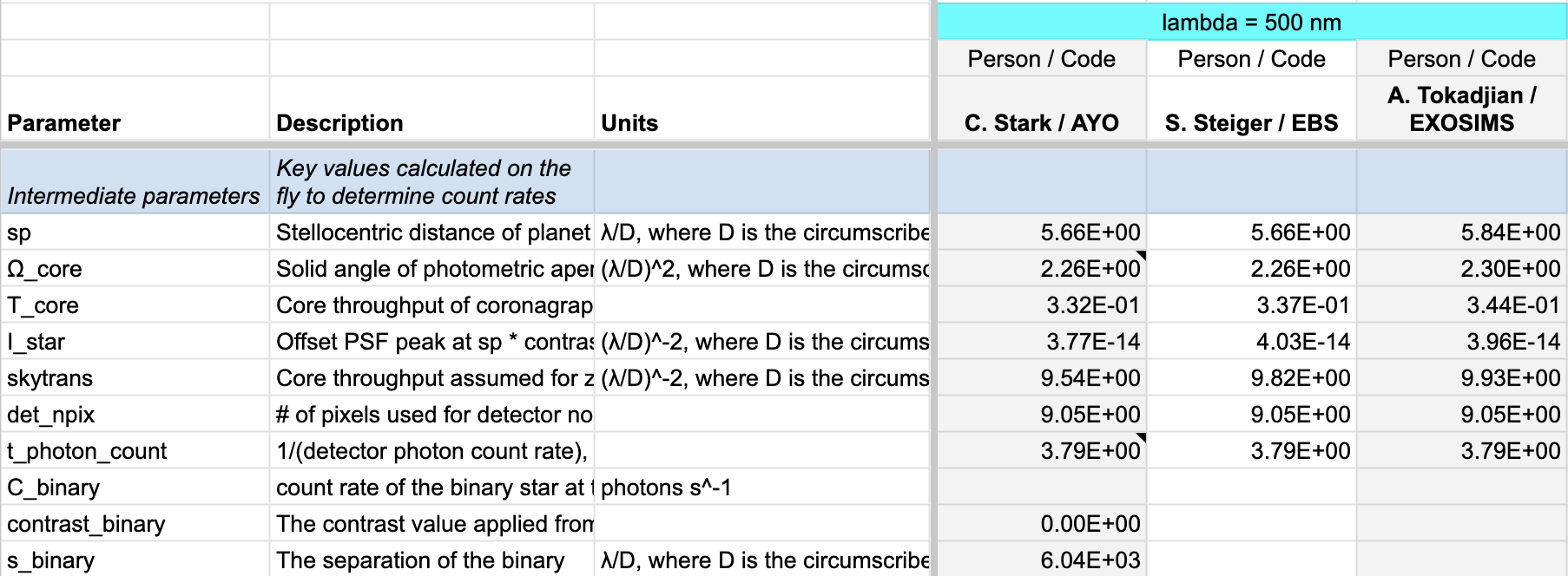}
    \caption{Same as Figure \ref{fig:spreadsheet_basic}, but listing the values of key intermediate parameters calculated by the ETCs.}
    \label{fig:spreadsheet_intermediate}
\end{figure}

\begin{figure}
    \centering
    \includegraphics[width=1.0\linewidth]{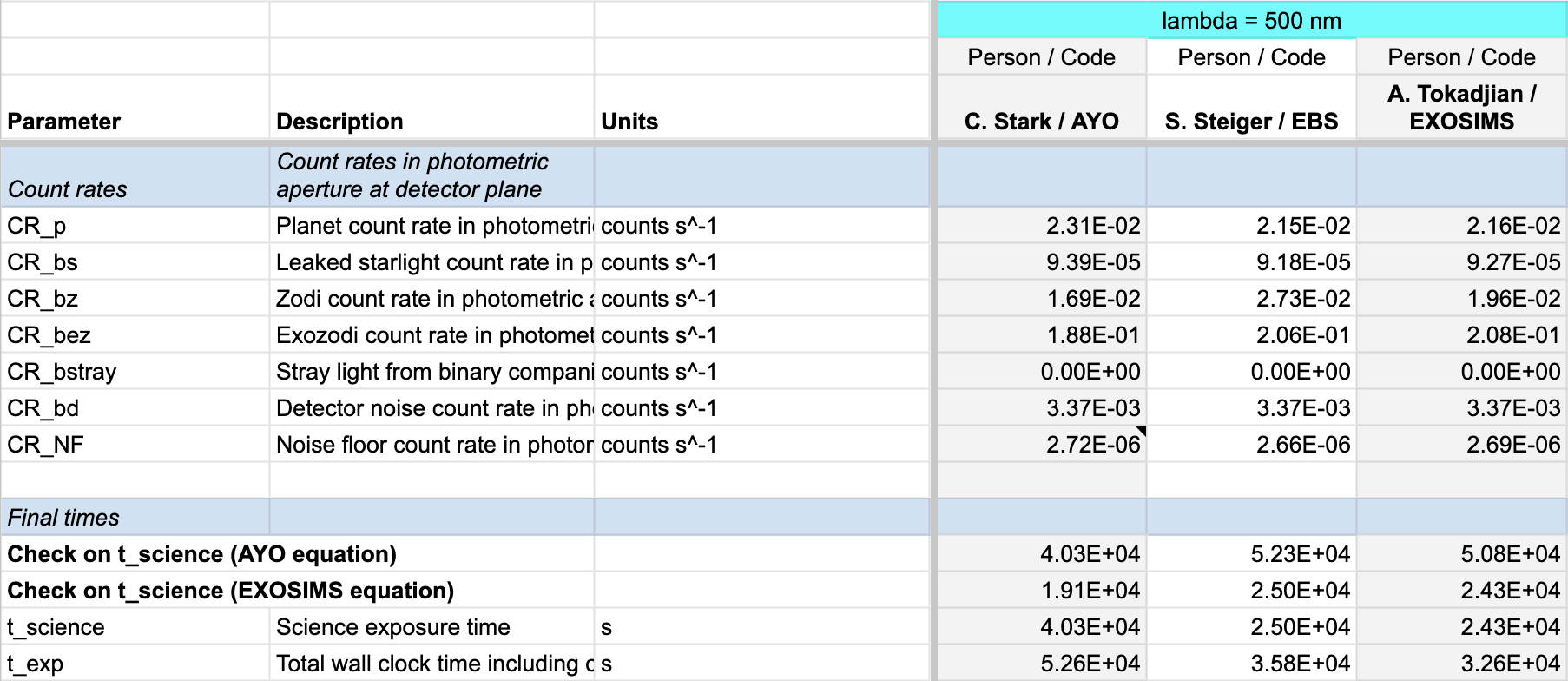}
    \caption{Same as Figure \ref{fig:spreadsheet_basic}, but listing the count rates of each source measured in the adopted photometric aperture at the detector plane, as well as the resulting exposure times. Rows in bold analytically determine the science exposure time using the listed count rates for both the AYO equation and the EXOSIMS equation and show that the differences in final science time are largely due to differences in the fundamental equations used.}
    \label{fig:spreadsheet_countrates}
\end{figure}

To calibrate the ETCs with one another, our Task Group created online spreadsheets for detection observations\footnote{https://www.starkspace.com/permanent/ETC\_cal\_detect.xlsx} and characterization observations\footnote{https://www.starkspace.com/permanent/ETC\_cal\_char.xlsx}. To ensure that the ETCs were calibrated over a broad range of scenarios, we examined five different fiducial stars, two observational scenarios (broadband detections and spectral characterizations), and two bandpasses for each scenario. Figures \ref{fig:spreadsheet_basic} through \ref{fig:spreadsheet_countrates} show the spreadsheet for HIP 32439; each of the five fiducial stars were tracked in a separate spreadsheet. The columns of this spreadsheet list the participants' names and codes that were used. Shown in Figures \ref{fig:spreadsheet_basic} through \ref{fig:spreadsheet_countrates} are the columns relevant to 500 nm detection observations. Additional columns were created for 1000 nm detection observations, and an additional spreadsheet was created for characterization observations to detect O$_2$ at 800 nm and H$_2$O at 1000 nm. 

The rows of the spreadsheet list various parameters, broken down by category. As shown in Figures \ref{fig:spreadsheet_basic} through \ref{fig:spreadsheet_countrates}, the ETC Calibration Task Group tracked a broad range of values that cover the entire process of exposure time calculation, ranging from basic inputs all the way to count rates and output exposure times. By comparing values at various stages of the calculations, we were able to identify the point at which codes disagreed and more easily identify the root causes of disagreement. As a policy, we chose to report values directly from the code as late in the process as possible, allowing us to explicitly check that the parameters had propagated through the code as expected.

The first category of the spreadsheet, shown in Figure \ref{fig:spreadsheet_basic}, lists ``basic parameters," which are fundamental inputs to exposure time calculations. Most of these parameters agree exactly, but not all. Notably, differences in the stellar input catalogs can be seen as small differences in stellar luminosity and distance. We also note that there appears to be $\sim10\%$ differences in flux zero point values as well as stellar magnitudes. However, this is in part due to how the codes determine stellar fluxes---whereas AYO adopts the apparent magnitudes from the Habitable Worlds Observatory's Preliminary Input Catalog \cite{tuchow2024} and interpolates Johnson flux zero points from Table A 2 of the Spitzer Telescope Handbook\footnote{https://irsa.ipac.caltech.edu/data/SPITZER/docs/spitzermission/missionoverview/\\spitzertelescopehandbook/18/}, EXOSIMS/EBS fits stellar spectral models to a broad range of photometry, then integrates the model flux over the desired bandpass (i.e., flux zero points are not directly used by EXOSIMS/EBS and the values reported here were ``backed out" of the calculations). EXOSIMS uses a combination of the stellar spectral models from the Pickles Atlas\footnote{https://www.stsci.edu/hst/instrumentation/reference-data-for-calibration-and-tools/astronomical-catalogs/pickles-atlas}\cite{pickles1998} and the  Bruzual-Persson-Gunn-Stryker Atlas\footnote{https://www.stsci.edu/hst/instrumentation/reference-data-for-calibration-and-tools/astronomical-catalogs/bruzual-persson-gunn-stryker-atlas-list}\cite{gunnstryker1983}.

There are also significant differences in solar longitude, the angle between the telescope-Sun vector and the telescope-target vector, which results from different levels of fidelity of the codes. Whereas the current version of AYO assumes all targets are observed at a solar longitude of 135$^{\circ}$, EBS assumes a pointing-independent zodiacal flux, and EXOSIMS adopts a user-supplied epoch of observation that determines the solar longitude. 

All of these differences in basic inputs will carry forward throughout the calculations, ultimately leading to differing exposure times. To help track this, we also reported the astrophysical fluxes received by the primary mirror, as shown in Figure \ref{fig:spreadsheet_astro}. Differences in the stellar input catalog, combined with the codes' different approaches to estimating stellar flux, leads to stellar fluxes that differ by $\sim10\%$. 

Differences in the fidelity of zodiacal light calculations as a function of telescope pointing can lead to differences approaching a factor of two. However, the differences in zodiacal light fluxes (due to differences in the solar longitude) will not significantly affect exposure times, as zodiacal flux is substantially less than exozodiacal flux in all cases.

Figure \ref{fig:spreadsheet_astro} shows only 6\% differences in exozodi flux that can easily be explained by minor differences in exozodiacal flux models. However, at a wavelength of 1000 nm we noticed large differences, approaching a factor of a few. This was determined to be due to a difference in how the codes modeled the color of exozodi, as discussed in Section \ref{sec:disagreement}.

Within the three ETCs we compared, many calculations are performed to convert the astrophysical fluxes shown in Figure \ref{fig:spreadsheet_astro} to exposure times. Our Task Group tracked these calculations by reporting out a set of ``intermediate parameters" calculated within the ETCs. Figure \ref{fig:spreadsheet_intermediate} shows this set of intermediate parameters. Many of these parameters are estimates of the coronagraph performance parameters at the location of the planet.

Small $\sim7\%$ differences can be seen in the estimates of the coronagraph core throughput, leaked starlight, and ``skytrans" values. As shown in Figure \ref{fig:spreadsheet_intermediate}, this can be explained in part by differences in the stellocentric distance of the planet, $s_{\rm p}$, which is due to differences in distance and luminosity of the star reported in the input stellar catalogs; small changes in the assumed angular separation of the planet can lead to perceived changes in coronagraph performance. 

However, even when these values are identical, differences $\sim$few percent remain. Values like the leaked starlight, $I_{\rm star}$, can vary rapidly with location in the image plane. These differences were determined to be largely due to differences in interpolation methods. Differences in spline or linear interpolation, as well as differences in the code base in which these standard interpolation methods are written, are responsible for the differences seen in Figure \ref{fig:spreadsheet_intermediate}. 

These differences were originally significantly larger. The Task Group identified that EXOSIMS/EBS was using 1D azimuthally averaged performance curves whereas AYO was estimating leaked starlight using 2D images, which can vary dramatically depending on the planet's precise location in two dimensions. For this calibration effort, the Task Group chose to adopt the azimuthally averaged performance curves for AYO as well, which is what is reported in Figure \ref{fig:spreadsheet_intermediate}.

Figure \ref{fig:spreadsheet_countrates} shows the final count rates for all sources, as measured at the detector plane within the adopted photometric aperture, as well as the resulting science exposure times (without overheads) and full exposure times (with overheads). The count rates listed include those of the exoplanet, the leaked starlight, zodiacal light, exozodiacal light, stray light due to binary companions, detector noise, and the noise floor term. HIP 32439 has a very distant binary companion at 79.2 arcseconds, which is beyond the limits of our extended PSF model and therefore has no contributed stray light.

AYO's 9\% larger stellar astrophysical flux results in planetary and stellar count rates that are both larger than those of EXOSIMS/EBS. Because the exozodiacal count rate for AYO is also marginally smaller that EXOSIMS, we'd expect the science exposure time to also be less. However, as shown in Figure \ref{fig:spreadsheet_countrates}, the science time for AYO is roughly 60\% longer than that of EXOSIMS/EBS. The reason for this is because of the fundamental differences of the equations used by these codes, as can be seen by comparing Equations \ref{eq:ayo_exposure_time} and \ref{eq:exosims_exposure_time}.

The bolded rows in Figure \ref{fig:spreadsheet_countrates} show an analytic check on the exposure times in which the count rates are plugged directly into each codes' respective exposure time equation (Equation \ref{eq:ayo_exposure_time} for AYO and Equation \ref{eq:exosims_exposure_time} for EXOSIMS/EBS). These bolded rows show that each code's science exposure time is correctly determined from the given count rates (e.g., the two cells with values of ``4.03E+04" show that the reported AYO science time is self-consistent with the reported count rates). These rows also show that if a code had adopted the other code's exposure time equation, the science times would agree to within $\sim30\%$ in this scenario.

\section{Results}

\subsection{Quantitative agreement of ETCs}
\label{sec:disagreement}

A summary of the agreement between the three codes studied is given in Table \ref{tab:count_rates}. We report the average percent disagreement for the three codes averaged over all wavelengths and targets studied for detection and characterization observations, broken down by count rates and science exposure times. For each count rate, ${\rm CR}$, the average disagreement is calculated as
\begin{equation}
    \frac{2}{3n_{\lambda}n_{\star}} \sum_{i=1}^{\rm n_\lambda} \sum_{j=1}^{\rm n_{\star}} \frac{{\rm CR}^{\rm AYO}_{i,j}\!\!-\!{\rm CR}^{\rm EBS}_{i,j}}{{\rm CR}^{\rm AYO}_{i,j}\!\!+\!{\rm CR}^{\rm EBS}_{i,j}} + \frac{{\rm CR}^{\rm AYO}_{i,j}\!\!-\!{\rm CR}^{\rm EXO}_{i,j}}{{\rm CR}^{\rm AYO}_{i,j}\!\!+\!{\rm CR}^{\rm EXO}_{i,j}} + \frac{{\rm CR}^{\rm EXO}_{i,j}\!\!-\!{\rm CR}^{\rm EBS}_{i,j}}{{\rm CR}^{\rm EXO}_{i,j}\!\!+\!{\rm CR}^{\rm EBS}_{i,j}},
\end{equation}
where $n_{\lambda}$ is the number of wavelengths and $n_{\star}$ is the number of stars. An identical formula was used for average disagreement of science exposure times, with science exposure time substituted for the count rate. This number provides a general estimate of the overall agreement. Further detail can be found in the online spreadsheet. 

The largest disagreement by source is zodiacal flux, with an average disagreement of $\sim100\%$. However, as previously noted, this is a minority contribution to the total count rate and therefore has a negligible impact on exposure times. The source of this disagreement is due to known differences in model fidelity and pointing assumptions. As such, this source of disagreement is not a primary concern. 

The second largest source of disagreement is exozodiacal flux. As mentioned in Section \ref{sec:cal_method}, the source of this disagreement is known to be primarily due to differences in assumptions about the color of exozodi. Whereas AYO assumes exozodi is the same color as the host star, EXOSIMS/EBS treats exozodi as an excess zodiacal light factor. As a result, the EXOSIMS/EBS exozodi flux had the same color as the Sun, not the target star. This issue in EXOSIMS/EBS was corrected after the conclusion of this comparison. If these updates reduced the disagreement in exozodi flux to zero, the average disagreement of the science exposure time would be reduced from 63\% to 40\%.

The disagreement in the planet count rates (${\rm CR}_p$) and stellar leakage count rates (${\rm CR}_{bs}$) are largely due to differences in the stellar flux, a result of using different stellar catalogs and different flux estimation methods. Adjusting these count rates by the ratio of stellar fluxes reduces the disagreement in ${\rm CR}_p$ and ${\rm CR}_{bs}$ to 5\% and 11\%, respectively. The remaining disagreement is due to differences in how the different input catalogs affect the planet's calculated separation as well as differences in coronagraph performance estimation. Adjusting count rates for both known differences in stellar flux modeling as well as exozodi modeling would reduce disagreement in science exposure time to 32\%. 

Detector noise count rates ($CR_{bd}$) are in near-perfect agreement at the $<5$\% level for both detection and characterization scenarios. Stray light count rates differ by $\sim30\%$, the cause of which was identified (see Section \ref{sec:stray_light}). The differences in noise floor count rates are similar to those of the stellar leakage, as expected given their proportionality.

\begin{table}[]
    \centering
    \begin{tabular}{l c c}
    Parameter & $\langle$Disagreement$\rangle$ & $\langle$Disagreement$\rangle$\\
            & (Detections) & (Characterizations) \\
    \hline
        CR$_p$ &  11\% & 18\%\\
        CR$_{bs}$ & 20\% & 26\%\\
        CR$_{bz}$ & 102\% & 125\%\\
        CR$_{bez}$ & 56\% & 74\%\\
        CR$_{bd}$ & 3\% & 5\%\\
        CR$_{bstray}$ & 30\% & 31\%\\
        CR$_{NF}$ & 20\% & 28\%\\
        $t_{exp}$ & 63\% & 60\%\\
    \end{tabular}
    \caption{Average disagreement for all of the studied count rates as well as the final science exposure time. This averages together all of the wavelengths and individual targets studied to give a general idea for the level of overall agreement between the three codes studied.}
    \label{tab:count_rates}
\end{table}

Table \ref{tab:count_rates} shows that the disagreement in science exposure times at the time of our analysis was $\sim60\%$. As discussed above, several updates to the code and stellar input catalogs could cut this disagreement in half. Below we discuss updates to the ETC codes that were identified as part of this effort.

\subsection{Updates to ETCs}
\label{sec:code_updates}

A number of differences between ETCs were discovered as a result of our calibration efforts. Some of these differences were intentional and due to different approaches. For example, EXOSIMS calculates the zodiacal flux at a given point in time assuming a given ecliptic latitude and solar longitude of the target at that epoch, while AYO accounts for the ecliptic latitude but ignores the time-varying solar longitude, and EBS prescribes 23 mag arcsec$^-2$ independent of pointing and wavelength---given the intended application and use of these different tools, none of these approaches are strictly correct or incorrect. While these intentional choices that result in disagreement were identified, they were not addressed by our group. However, our efforts did find several issues in AYO, EXOSIMS, and EBS that could be addressed to improve agreement.

\subsubsection{Exozodi color}
Our calibration efforts showed that while AYO, EXOSIMS, and EBS roughly agreed on the exozodiacal astrophysical flux at 500 nm, they disagreed on the flux at 1000 nm by factors of a few. This disagreement also varied with spectral type. Our group identified the cause of this disagreement: a difference in how AYO, EXOSIMS, and EBS treat exozodi color. EXOSIMS adopts an exozodi color identical to that of our solar system's zodi, whereas AYO adopts an exozodi color identical to that of the target star.

The color of the zodiacal light is, to first order, the same as that of the Sun\cite{leinert1998}. In other words, zodiacal dust is a gray scatterer. While we have very limited knowledge of the color of warm \emph{exozodiacal} dust, there is no basis to assume that on average it is not also a gray scatterer. Our task group agreed that exozodiacal dust models in ETCs should follow the color of the target star. Since the conclusion of our analysis, a pending update for EXOSIMS has been developed to calculate exozodiacal color in the manner described here.

\subsubsection{Stray light from binary stars}
\label{sec:stray_light}
Our calibration efforts also revealed major differences in the leaked stray light from binary stars. The count rates between EXOSIMS and AYO differed by roughly two orders of magnitude, with AYO predicting higher count rates. Our task group consulted with multiple experts on coronagraph simulation and stray light and concluded that neither code was properly calculating the count rate from binary stray light. Ultimately the correct approach was determined and the proper equations were derived (see Equation \ref{eq:AYO_stray_light}). The resulting values lie between the initial AYO and EXOSIMS values. The AYO calculations have been updated to use the new expression. A problem report has been filed for the EXOSIMS software project and an update is forthcoming.

\subsubsection{Stellar input catalogs}
Many disagreements, including stellar fluxes, planet count rates, and estimated coronagraph performance, can be explained by differences in the input stellar catalog. AYO uses the Habitable Worlds Observatory's Preliminary Input Catalog (HPIC)\cite{tuchow2024}, the branch of EXOSIMS used in this comparison uses ExoCAT1\cite{turnbull2015}, and EBS uses the NASA ExEP HWO Target List\cite{exeptargetlist}. These catalogs were developed using different sources for distances, magnitudes, etc. The differences in the choice of input stellar catalog are partly due to code heritage, but also partly due to functional requirements--some codes require different sets of input resulting from differences in methodologies. 

Differences in this input catalog can produce substantial differences in the estimated exposure time, especially if the planet is near the coronagraph's inner working angle (IWA). Near the IWA, the contrast and throughput can be a strong function of separation from the star. As such, small differences in assumed luminosity and distance of the star, which affect the apparent separation of the planet, can affect the estimated coronagraph performance. Our task group finds that the community could help retire uncertainty in these calculations by creating a single stellar input catalog that provides all ETC codes with all necessary inputs.

\subsection{Limitations of this comparison}

This effort attempted to perform an exhaustive comparison of three ETCs. However, due to the sheer number of parameters governing exposure time calculations, not all parameters could be investigated independently. Several limitations of our comparison are noteworthy.

First, we considered only a single coronagraph model, an optical charge 6 vortex coronagraph designed for the off-axis segmented USORT aperture. This choice turned out to be fortuitous, as it helped highlight the importance of a careful distinction between inscribed and circumscribed telescope diameter (see discussion in Section \ref{sec:lessons}). Substantial effort was devoted during our comparison to ensuring a proper treatment of the correct telescope diameter throughout the codes. However, by not considering additional coronagraphs that stressed other parameters, like outer working angle, throughput, etc., it is possible that additional issues concerning how the codes estimate coronagraph performance were left uncovered.

All three ETCs included in this effort address stray light in  a limited sense. Only the diffracted stray light at large separations from binary companion stars, due to high frequency artifacts like contamination or surface roughness, are included. Additional effects, like stray light from background objects and incoherent stray light from the target star due to near angle scatter, may also play an important role. These effects need to be addressed in the coming years as stray light is expected to play an important role in exoEarth detection and characterization at the 10$^{-10}$ level.  

Both AYO and EXOSIMS incorporate a pitch angle and pointing of the observatory when calculating local zodi contribution. However, EBS zodi flux does not depend on pointing but uses the canonical 23 mags arcsec$^{-2}$ for all systems and bandpasses. As such, we could not define a common pointing assumption to accurately compare the zodiacal light calculations of all three codes. Since local zodi flux is small compared to other sources of noise, this difference has a negligible effect on our exposure time calculations.


Our comparison was also limited in scope to the five fiducial stars chosen for analysis: one F star, two G stars, and two K stars. Although these types make up the majority of the current HWO target list\cite{exeptargetlist}, it was not possible to consider all types and distances. Notably A stars at large distances and nearby M stars were not included. Since many important parameters depend on stellar type, such as the stellocentric distance of the planet, extending the spectral type could highlight additional sources of disagreement. Furthermore, HIP 77052 was the only star included with a binary companion that contributed stray light. 

\section{Lessons learned}
\label{sec:lessons}

The ESYWG's ETC Calibration Task Group found the calibration effort to be valuable and illuminating. Coronagraphic exposure time calculations are complex, can be affected significantly by many parameters, and require a detailed understanding of the astrophysical assumptions, coronagraphy, and the numerical innerworkings and methodology of the code. As such, an intimate familiarity with the code is essential. By conducting this effort, the developers of EXOSIMS, AYO, and EBS all became more familiar with the underlying calculations. 

By the same token, the complexity of such calculations provides many opportunities for error. This effort helped to highlight common user errors, errors in inputs, and errors in methods.

\subsection{Exposure time alone is inadequate to compare ETCs}

We find that when validating coronagraphic ETCs, comparing exposure time alone is inadequate. Given the complexity of coronagraphic exposure time calculations and the large number of parameters involved, it is possible to agree on exposure times while disagreeing on fundamental terms that govern exposure time. It is therefore essential to compare critical values like astrophysical fluxes and count rates broken down by source to uncover all possible sources of disagreement within ETCs. Furthermore, breaking the calculations down into these critical terms is required to understand the cause for any disagreement in exposure times.

\subsection{Codes should be validated using multiple and varied observing scenarios}

Some of the discrepancies uncovered as a part of this effort were only made clear through the analysis of one or a few of the 20 total observing scenarios studied (5 stars, 2 wavelengths, 2 modes -- detection and characterization). For example, only one of the targets studied (HIP 77052) had a close enough companion to be able to test and compare the stray light modules used in each of the codes. In doing so, the major discrepancies described in Section \ref{sec:stray_light} were able to be found.

Additionally, testing multiple wavelengths was essential to be able to uncover the different ways the codes handled exozodi color and also helped to debug other sources of error such as differences in stellar input catalogs. One example was a scenario in which one catalog yielded different I-band photometry as compared to another even though their V-band photometry was in perfect agreement, resulting in different stellar fluxes at 1000 nm.  

\subsection{Small differences in inputs can lead to large differences in outputs}

The coronagraphic ETCs rely on a long list of inputs, all of which are subject to differences. As an example, we consider differences in stellar parameters that flow down to all aspects of the ETC.

All three ETCs adopted slightly different stellar catalogs, for which photometry differed from star to star. Differing or missing photometry resulted in astrophysical fluxes that differed by a few percent. In addition to that, the codes handled estimating stellar fluxes differently. Whereas AYO relied on interpolation of measured photometry only, EXOSIMS fit stellar model templates to existing photometry and used it to estimate the flux over any given bandpass---these differences resulted in additional differences at the few percent level.

Further, discrepancies in stellar distances and luminosities resulted in differences in the projected separation of a planet at the EEID. When near the inner working angle of the coronagraph, these small differences in the planet's projected separation can mean large differences in the contrast of the suppressed star, which in some cases could result in large differences in exposure times. In other words, two users can have identical ETCs and arrive at very different exposure times if their input catalogs differ. The Task Group concluded that the best way to address this is to track the progress of an exposure time calculation from inputs to exposure times, as was enabled by our spreadsheet analysis.

As another example of how differences in input assumptions can lead to problematic results, we note that not all of the fiducial stars are plausible targets for all types of observations that we considered. For example, the detection time at 500 nm for HIP 32439 is $\sim5$ hours, but the characterization time to search for O$_2$ is an astounding $\sim200$ days. This highlights that not all fiducial stars make good targets for all coronagraphs and a mission's target list must be optimized for the coronagraph design and scientific requirements.  

\subsection{Coronagraphic ETC workshops are valuable}

Complex codes require experienced users to make sure that all inputs are specified properly. To this end, workshops that educate users and train the community to use these codes are highly valuable. Several recent community workshops on yield calculations covered the topic of ETCs--valuable resources from these workshops can be found online, including recorded talks, slide decks, and hands-on workbooks\footnote{2024 Sagan Summer Workshop: https://nexsci.caltech.edu/workshop/2024/}\footnote{Exoplanet Yield Modeling Tools Workshop: https://exoplanets.nasa.gov/exep/events/490/exoplanet-yield-modeling-tools-workshop-remix/}. Future workshops focused on coronagraphic exposure time calculations would continue to benefit the community.

\subsection{Code development should enable cross-model validation}

Throughout this effort, the importance of having easy hooks in the code to be able to report intermediate variables and check calculations was made obvious. Current and future ETC developers should make it a priority to not obfuscate access to these variables either through careful logging or by making the code base modular enough that these values can be checked at many stages of a calculation. In some instances this could (or should) include plots to visualize dependent variables. An example relevant to this work would be to plot the coronagraph throughput as a function of separation.

\subsection{Differences in methods exist that can be addressed}

The ETC Calibration Task Group noted some differences in methods between codes that could be updated in the future. Most of these issues are not expected to dominate differences in exposure times and are the result of intentional choices made given the limitations/approach of the codes. These issues were not addressed in this study, but rather mitigated by performing manual checks on count rates and exposure times. However, the differences noted below likely could be addressed in the future:
\begin{itemize}
    \item Zodi color: EXOSIMS and AYO adopt zodiacal color variations based on \cite{leinert1998}, while EBS adopts 23 mag arcsec$^{-2}$ at all wavelengths. While the zodi fluxes reported by these codes roughly agree at V band, at 1000 nm EBS underpredicts zodi flux by a factor of $\sim$3.
    \item Zodi time dependence: EXOSIMS calculates zodi flux as a function of ecliptic latitude and solar longitude for a given epoch ignoring seasonal zodi variations, whereas AYO does not attempt to schedule observations and therefore only accounts for the ecliptic latitude, while EBS does not attempt to take pointing variation into account at all. Ideally all codes would calculate zodi flux as a function of time including both the pointing of the target as well as seasonal zodi variations.
    \item Differing definitions of the PSF core: EXOSIMS adopts a circular photometric aperture of fixed size (selected by radius), whereas AYO adopts a core that can vary in shape based on the ratio of the PSF to its peak (selected based on intensity of the PSF) and optimally selects the PSF core ratio that minimizes exposure times. The PSFs of some coronagraphs can be significantly distorted near the inner working angle. As such, ideally all codes would adopt a PSF core defined in terms of normalized PSF intensity instead of radius.
    \item Differing exposure time equation formats: AYO includes the planet's count rates as a noise source for both detections and characterizations, whereas EXOSIMS and EBS include it only for spectral characterizations as suggested by \cite{nemati2023}. These differences are only expected to affect the brightest, nearest stars, for which exposure times are relatively short and therefore will not affect exoplanet yields. However, existing coronagraphic data sets could be used to inform which approach should be adopted.
    \item Inscribed diameter vs. circumscribed diameter: for a segmented aperture with a jagged outer edge, there is no clear definition of the telescope's diameter. This can cause confusion when working in units of $\lambda/D$, leading to instances in which the coronagraph's performance is applied in the wrong units. Because many coronagraph designs adopt pupil stops or Lyot stops that are approximately limited to the inscribed diameter of the telescope, one may prefer to work in units of $\lambda/D_{\rm ins}$ (some subroutines of EXOSIMS adopt the inscribed diameter). However, not all starlight suppression systems are limited in this way. As a result, our Task Group concluded that the preferred units for $\lambda/D$ adopt the circumscribed diameter of the telescope for $D$.  The choice of circumscribed diameter is consistent with the yield input package standard interface and the automated CDS software pipeline\cite{CDS}. 
\end{itemize}

\subsection{Differences in methods exist with no easy fixes}
Some differences in calculations were simply the result of intentional differences in methods, with no clear solution. Again, these differences were mitigated by performing manual checks in the comparison spreadsheet. Here are several examples:
\begin{itemize}
    \item Differing ``bookkeeping" methods: some ETCs keep track of certain parameters separately while others lump them into a single term. The reasons for this can range from differing conventions to choices made for the purpose of efficient computation, and correcting this could require major rework of established codes with little functional benefit. These differences make cross-model validations more challenging.
    \item Differing interpolation methods: many aspects of exposure time calculation require interpolation, e.g., when estimating the contrast at a given separation for a star of a given diameter. While differences in interpolation methods (e.g., linear vs. quadratic vs. spline, linear-space vs. log-space, etc.) usually have small effects on the result, in some cases they can have significant effects (e.g., the raw contrast near the inner working angle). Perfect agreement on all interpolations would be both time-consuming and challenging given the various coding languages involved.
    \item Differing methods for calculating core-integrated values: when calculating exposure times, the count rates must be integrated over the photometric aperture. Oftentimes shortcuts are taken for numerical expediency (e.g., the count rate per pixel is evaluated at the center of the core and then multiplied by the number of pixels in the core). These shortcuts, and differences between the implementation of the shortcuts, can result in differences in exposure times. Adopting identical methods could require some codes to run much more slowly.
    \item The systematic noise floor, implemented as ${\rm CR_{nf}}$ in Equation \ref{eq:ayo_exposure_time}, ultimately sets the detection and characterization limits. All three ETCs examined during this study can adopt a uniform noise floor independent of separation (appropriate if the telescope stability sets the noise floor independent of raw contrast) or a noise floor that is proportional to the raw contrast (i.e., a ``post-processing factor"). However, it is unclear at this point what the correct approach is for HWO. As such we find that ETCs should clearly report their approach to estimating the systematic noise floor.
\end{itemize}

\section{Conclusions}
The ESYWG ETC Task Group has concluded its cross-model validation of three existing ETCs. We find that the ETCs use a broad variety of differing methods, assumptions, and inputs that produce variation in the final exposure times at the $\sim60\%$ level. The causes for the disagreement have largely been identified, flagged for further development efforts, and in some cases retired since the conclusion of this effort. We expect that addressing the flagged efforts will bring the ETCs to within better than $\sim30\%$ agreement.

The ETC Task Group has developed an ETC comparison spreadsheet that allows users to compare inputs, astrophysical fluxes, intermediate parameters, count rates, and exposure times. By doing so, users can more easily track and identify the sources of disagreements. Future efforts to validate ETCs (e.g., after additional development) could take advantage of this framework.

\printbibliography

\end{document}